\documentstyle[amssymb,aps,preprint,prl]{revtex}
\draft

\begin{document}
\title{$V-I$ characteristics in the vicinity of order-disorder transition in vortex matter}

\author{Y. Paltiel$^1$, Y. Myasoedov$^1$, E. Zeldov$^1$, G. Jung$^{1,2,\dag}$,
M. L. Rappaport$^1$, D. E. Feldman$^{1,\ddag}$, M. J. Higgins$^3$,
and S. Bhattacharya$^{3,4}$}

\address{$^1$Department of Condensed Matter Physics, Weizmann Institute of
Science, Rehovot 76100, Israel}
\address{$^2$Department of Physics, Ben-Gurion University of the Negev, Beer-
Sheva 84105, Israel}
\address{$^3$NEC Research Institute, 4 Independence Way, Princeton, New Jersey
08540}
\address{$^4$Tata Institute of Fundamental Research, Mumbai-400005, India}

\date{\today}
\maketitle

\begin{abstract}

The shape of the $V-I$ characteristics leading to a peak in the differential
resistance $r_d=dV/dI$ in the vicinity of the order-disorder transition in NbSe$_2$
is investigated. $r_d$ is large when measured by dc current. However, for a small
$I_{ac}$ on a dc bias $r_d$ {\em decreases} rapidly with frequency, even at a few
Hz, and displays a large out-of-phase signal. In contrast, the ac response {\em
increases} with frequency in the absence of dc bias. These surprisingly opposite
phenomena and the peak in $r_d$ are shown to result from a dynamic coexistence of
two vortex matter phases rather than from the commonly assumed plastic depinning.

\end{abstract}

\pacs{PACS numbers: 74.60.Ec, 74.60.Ge, 74.60.Jg}


Measurement of $V-I$ characteristics or of the differential resistance $r_d=dV/dI$
in superconductors is a very common method to investigate vortex dynamics and to
identify various possible pinned and moving phases of the vortex matter. In
particular, the specific shape of $V-I$ that leads to a peak in $dV/dI$ has
attracted much attention
\cite{won,shoboPRL,shobo,garten,hellerqvist,geers-kes,xiao00,kv,ryu,olson,marchetti}.
In numerical simulations this peak in $r_d$ is usually ascribed to plastic vortex
depinning followed by dynamic ordering of the lattice
\cite{kv,ryu,olson,marchetti}. However, magnetic decorations and SANS studies,
which have observed the plastic deformation of the lattice near the depinning
followed by the dynamic ordering \cite{yaron,duarte,pardo}, did not find the
predicted peak in $dV/dI$ \cite{yaron,duarte}. In addition, in clean systems like
NbSe$_2$ the peak in $dV/dI$ is present surprisingly only in the lower part of the
peak effect (PE), whereas in the rest of the $H-T$ phase diagram the $V-I$ curves
are concave upward with no peak in $dV/dI$ \cite{shoboPRL,shobo,evacut}. Moreover,
in the same region where the peak in $r_d$ is observed, a number of anomalous
vortex matter properties were recently found, including the striking observation
that for an ac current the apparent vortex mobility {\em increases} rapidly with
frequency\cite{hend98,yosna,metlushko,gordeev,kwok}. Several ideas and models have
been proposed in which the ac agitation facilitates the plastic vortex depinning.
Yet another important paradox has received little attention: when the ac current is
superposed on a dc bias the opposite behavior is observed, i.e., the apparent
vortex mobility {\em decreases} with frequency, even at frequencies as low as
several Hz \cite{shobo}. None of the models that describe the peak in $r_d$ or the
mobility enhancement with frequency \cite{kv,ryu,olson,marchetti,hend98} has
resolved this apparent paradox.

In this paper we demonstrate that the shape of the $V-I$ curves, the peak in
$dV/dI$, and the opposite frequency dependencies stem, instead, from a dynamic
coexistence of two vortex matter phases. A highly-pinned metastable disordered
phase (DP), generated at the sample edges, anneals into a weakly-pinned equilibrium
ordered phase (OP) in the bulk of the sample \cite{yosna,marchevsky,giller}. The
specific shape of the dc $V-I$ curves results from the fact that most of the sample
is in the metastable DP at low currents, but in the OP at high currents. This
dynamic transformation results in a peak in $dV/dI$. If measured by a small ac
current superposed on a dc bias, we find that the peak in $dV/dI$ decreases with
frequency and displays a unique out-of-phase signal due to the slow transformation
process. In contrast, for an ac current with no dc bias only the edges of the
sample are contaminated by the DP, resulting in the opposite behavior, i.e., the
voltage response grows with the ac frequency.

Transport measurements were carried out on several Fe-doped (200 ppm) NbSe$_2$
crystals in strip-like four-probe configuration in applied field $H \|$ c-axis. The
data presented here are for a crystal $2 \times 0.4 \times 0.04$ mm$^{3}$ with
$T_c=5.6$ K, $H_{c2}(4.2 \rm K)=1$ T, and the PE field $H_{p}(4.2 \rm K)= 0.53$ T.
Very low contact resistance of $\sim 10$ m$\Omega$ was achieved with large current
contacts of Au evaporated onto a freshly cleaved surface. By immersing the crystals
in liquid He, currents up to 100 mA could be applied with negligible heating.
Square wave or sinusoidal $I_{ac}$ was used and the corresponding $V_{ac}$ was
measured by a lock-in amplifier.

Figure 1 shows $V_{ac}$ vs. $I_{ac}$ and $V_{dc}$ vs. $I_{dc}$ in the vicinity of
the order-disorder transition in the lower part of the PE at $H=0.44$ T. The dc
curve (solid circles) starts to increase in \emph{concave} form above 18 mA and
then rapidly turns \emph{convex}. At higher currents linear flux flow behavior is
obtained. Although the form of the dc $V-I$ curve may seem rather conventional, we
find that in NbSe$_2$ the convex shape is present only in the lower part of the PE,
while in the rest of the phase diagram the curves are always concave upward,
consistent with previous reports \cite{shoboPRL,shobo,evacut}. $V_{ac}$ vs.
$I_{ac}$ measured at various frequencies in Fig. 1 are remarkably different from
the dc $V-I$. Even at a frequency as low as 1 Hz the apparent $I_c$ is much lower
and the voltage response below 20 mA is strongly enhanced. Furthermore, the
apparent vortex mobility increases rapidly with frequency, as noted previously
\cite{hend98,yosna,metlushko}.

Figure 2 shows the differential resistance $r_d$ measured by superimposing a small
$I_{ac}$ (0.1 to 1 mA) on $I_{dc}$, $r_d=V_{ac}/I_{ac}$, along with $r_d=dV/dI$
obtained by numerical differentiation of the dc $V-I$ at $H=0.31$ T. At this
slightly lower field within the PE the dc $V-I$ characteristic (solid curve) is
more gradual and turns convex above $\sim 20$ mA. At the inflection point, $dV/dI$
displays a large peak reaching three times the flux flow $r_f$. The remarkable
result here is that the ac $r_d$ is significantly different from the dc value
\cite{shobo} and it decreases rapidly with ac frequency. Note that the peak in ac
$r_d$ is suppressed to about half of the dc value already at $f$ as low as 3 Hz,
indicating the existence of very long characteristic timescales that are even
longer than the vortex transit time across the sample, as described below. None of
the microscopic bulk mechanisms, such as plastic depinning or dynamic ordering, can
account for such long timescales. Moreover, due to the high sensitivity of lock-in
technique, the ac $r_d$ vs. $I_{dc}$ is commonly integrated to derive the full
$V-I$ characteristics \cite{garten,hellerqvist}. The important conclusion from Fig.
2 is that the integral of the ac $r_d$ is {\em not equal} to the dc $V-I$.

Figures 1 and 2 display a striking qualitative difference: for a pure $I_{ac}$ the
measured $V_{ac}$ {\em increases} with frequency, whereas for a small $I_{ac}$
superposed on a larger $I_{dc}$ the resulting $V_{ac}$ {\em decreases} with $f$.
The lower panel of Fig. 2 shows another puzzling aspect of the data, which is a
large out-of-phase component that appears only in the non-linear part of the $V-I$
and is absent in the flux-flow region. An out-of-phase signal is a common feature
in ac susceptibility measurements, where the amplitude of the current induced in
the sample and the dissipation level depend on the excitation frequency. In
transport measurements, in contrast, the amplitude of the current is fixed by the
external circuitry and therefore the voltage response, as a rule, is frequency
independent and does not show any out-of-phase signal at low frequencies. To the
best of our knowledge this is the first published report of an imaginary $r_d$,
which further emphasizes the anomalous vortex dynamics in the lower part of the PE.
Note that the out-of-phase $r_d$ is non-monotonic with frequency: it vanishes in
the limit of high and low $f$ and is largest for the 22 Hz data.

We now discuss the results in view of the recent understanding that the PE reflects
a disorder-driven first-order phase transition from a weakly pinned OP (Bragg
glass) with a low critical current density $J_c^{ord}$ into a strongly pinned DP
with a high $J_c^{dis}$ \cite{yosprl,nurit,ravikumar}. Local measurements have
demonstrated that below the transition, in the lower part of the PE, in the
presence of transport current a supercooled metastable DP is formed at the sample
edge because of non-uniform vortex penetration through the surface barriers
\cite{yosna,marchevsky,giller}. As the vortex lattice moves across the sample, the
metastable DP with a high concentration of dislocations gradually anneals into the
dislocation-free OP. We describe the annealing stage of the DP by its local
critical current density $J_c(x)$ which has a non-equilibrium excess value
$\widetilde{J_c}(x)=J_c(x)-J_c^{ord}$ relative to the fully annealed OP. Since in
low-temperature superconductors thermal activation is negligible, the sole
annealing mechanism of the metastable DP is through a current-driven displacement
that allows rearrangement and disentanglement of the vortices during the motion. We
therefore assume, for simplicity, that the relative annealing of $\widetilde{J_c}$
upon displacement by a small $\Delta x$ is given by $\Delta x/L_r$, where $L_r$ is
a characteristic relaxation length over which the DP anneals into the OP. Since the
lattice flows with velocity $v$, $\widetilde{J_c}$ at $x+\Delta x$ and at time
$t+\Delta t=t+\Delta x/v$ is thus described by $\widetilde{J_c}(x+\Delta x,
t+\Delta x/v) =\widetilde{J_c}(x,t)(1-\Delta x/L_r)$, which leads to the partial
differential equation of the annealing process
\begin{equation}\label{1}
  \partial \widetilde{J_c}(x,t)/\partial x +
  (1/v)\partial \widetilde{J_c}(x,t)/\partial t = -
  \widetilde{J_c}(x,t)/L_r(v),
\end{equation}
with a boundary condition at $x=0$, where vortices penetrate into the sample, of
$\widetilde{J_c}(0,t)=J_c^{dis}-J_c^{ord}$. A key aspect of the annealing process
is that the relaxation length $L_r$ crucially depends on the displacement velocity
$v$. Fast transient measurements \cite{hend96} show that at low velocities $L_r$ is
large, whereas for a potential landscape that is strongly tilted by a large driving
force the disentanglement is very rapid, so that empirically $L_r\simeq
L_0(v_0/v)^\eta=L_0(V_0/V)^\eta$. Here $\eta$ is typically in the range of 1 to 3,
$L_0$, $v_0$, and $V_0$ are scaling parameters, $V=vBl$ is the measured voltage
drop, $B$ is the magnetic field, and $l$ is the distance between the voltage
contacts. We now demonstrate that these simple assumptions describe all the
essential experimental observations.

We first analyze the time independent behavior. The dc solution of Eq. \ref{1} is
$J_c^{dc}(x)= (J_{c}^{dis}-J_{c}^{ord})\exp (-x/L_r)+J_c^{ord}$, as shown
schematically in Fig. 3 inset. By integrating over the width $W$ we obtain the
total critical current of the sample $I_{c}=d \int_0^{W}J_c^{dc}(x)dx$:
\begin{equation}\label{2}
  I_{c}(L_r)=
(J_{c}^{dis}- J_{c}^{ord})[1-e^{-\frac{W}{L_r(V)}}]L_r(V)d +
I_{c}^{ord},
\end{equation}
where $I_c^{ord}=J_c^{ord}Wd$ and $d$ is the sample thickness. Note that $I_c$
depends on $L_r$, which in turn depends on voltage $V$. This property is central to
the described phenomena: Coexistence of the DP and OP results in an inhomogeneous
sample and moreover the degree of the inhomogeneity, $J_c(x)$, changes with vortex
velocity. As a result, the total $I_c$ of the sample is not fixed, but rather
changes with voltage.

The dc $V-I$ characteristics can be derived as following. We write for simplicity
the $V-I$ of the OP as $V=r_f(I-I_c^{ord})$, where $r_f$ is the flux-flow
resistance. Similarly, when the entire sample is in the DP, $V=r_f(I-I_c^{dis})$
with $I_c^{dis}=J_c^{dis}Wd$. These two asymptotic $V-I$ solutions are shown by the
dashed lines in Fig. 3. Here we have assumed for simplicity that the flux-flow
resistance $r_f$ is the same for any of the phases. As a result, when the two
phases coexist $V=r_f(I-I_c)$, where $I_c$ is given by Eq. \ref{2} and is voltage
dependent through $L_r(V)$. Hence an analytical $I(V)$ relation can be written
directly as $I= V/r_f + I_c(L_r(V))$, and the resulting non-linear $V-I$
characteristic is shown by the solid curve in Fig. 3. At very low voltages $L_r$ is
larger than the sample width, namely the entire sample is contaminated by the DP,
and hence the $V-I$ initially follows the asymptotic dashed line of the DP with
$I_c= I_c^{dis}$. At high vortex velocities $L_r$ becomes very short, most of the
sample is in the OP and the $V-I$ approaches the asymptotic line of the OP with
$I_c= I_c^{ord}$. In the crossover region a specific shape of the curve with an
inflection point is obtained alike the experimental dc $V-I$ curves in Figs. 1 and
2. This shape is the result of a continuous decrease of the total $I_c$ of the
sample from $I_c=I_c^{dis}$ to $I_c=I_c^{ord}$ with increasing current. The exact
curvature in the crossover region depends on the parameters $\eta$, $L_0$, and
$V_0$ (see Fig. 3 caption) and may be either gradual or very steep, and may even
obtain a negative slope resulting in an S-shaped characteristic as found recently
in the lower part of the PE \cite{zhukov}.

In order to understand the frequency dependence of the differential resistance
shown in Fig. 2 and in more detail in Fig. 4a, we solve Eq. \ref{1} for a small
periodic velocity perturbation $v=v_{dc}+v_{ac}e^{i\omega t}$ caused by applied
current $I_{dc}+I_{ac}e^{i\omega t}$. In this case
$L_r(v)=L_{dc}+(dL_r/dv)v_{ac}e^{i\omega t}$. Following a simple calculation and
keeping only the linear terms in $v_{ac}$, we obtain
$J_c(x,t)=J_c^{dc}(x)+J_c^{ac}(x)e^{i\omega t}$, where
$J_c^{ac}(x)=(J_c^{dis}-J_c^{ord})(dL_r/dv)(v_{dc}v_{ac}/L_{dc}^2)
e^{-x/L_{dc}}(1-e^{-i\omega x/v_{dc}})/i\omega$. Note that $J_c^{ac}(x)$ is
negative because of a negative $dL_r/dv$, reflecting the fact that the system
becomes more ordered with increasing $v$. In the limit of low frequency, $J_c(x,t)$
slowly varies between two extreme dc solutions determined by $L_{dc}\pm
(dL_r/dv)v_{ac}$, as shown in Fig. 3 inset. At the minimum value of the current,
$I_{dc}-I_{ac}$, the vortex velocity is lowest, $L_r$ is largest, and hence
$J_c(x)$ is highest. At the maximum of the current, $I_{dc}+I_{ac}$, $L_r$ is
smallest and the sample is `cleanest', resulting in a large enhancement in vortex
velocity. Hence the large $r_d$ at low $f$ arises from the varying contamination of
the sample which significantly amplifies the voltage response. However,
modification of the sample contamination is a slow process because the only
mechanism of enhancement of the local disorder in the bulk is by transporting a
more disordered lattice from the edge of the sample. This results in characteristic
timescales comparable with the vortex transit time across the sample $\tau_t=W/v$
(or even longer, see below).

To obtain the full frequency dependence of $r_d$ we calculate $I_{c}^{ac}=d
\int_0^WJ_c^{ac}(x)dx$, and by noting that $V_{ac}=r_f(I_{ac}-I_c^{ac})$, we find
$r_d=V_{ac}/I_{ac}= (1/r_f+I_c^{ac}/V_{ac})^{-1}$. This result shows that the
enhancement of $r_d$ results from the fact that $I_c^{ac}$ is negative, and
depending on the $L_r(v)$ parameters, $r_d$ can become infinite and even negative
for the case of S-shaped $V-I$ curves. The full analytical expression for $r_d$ is
too extensive to be presented here. Figure 4b therefore shows the calculated real
and imaginary parts of $r_d/r_f$ vs. frequency at the operating point in Fig. 3. At
low frequencies the in-phase $r_d$ is large and it decreases towards $r_f$ when $f$
approaches the transit frequency $f_t=1/\tau_t$. We can understand this behavior as
follows. As indicated by the arrows in Fig. 3, at low frequencies $r_d$ is given by
$dV/dI$ which can be very large depending on the specific shape of the dc $V-I$
curve. However, the surprising result here is that at higher frequencies, in
contrast to the common belief, the experimental $r_d=V_{ac}/I_{ac}$ does not
measure the true $dV/dI$. At high $f$ the local degree of disorder $J_c(x,t)$
cannot adjust to the rapid variations and therefore $J_c(x,t)$ is fixed at
$J_c^{dc}(x)$, and $J_c^{ac}(x)$ and $I_c^{ac}$ vanish. As a result, at high
frequency the ac signal, instead of following the dc curve, follows a trajectory
shown by the dotted line in Fig. 3. This line is the $V-I$ characteristic of a
sample with a fixed $I_c=I_c^{dc}$, $V=r_f(I-I_c^{dc})$, resulting in $r_d=r_f$, as
indeed observed experimentally in Figs. 2 and 4a.

The rearrangement of the bulk disorder always lags the external ac drive, giving
rise to a pronounced imaginary $r_d$ component (Fig. 4b) that has a maximum at
intermediate frequencies. The qualitative agreement between the results of our
simplified calculations in Fig. 4b and the experimental data in Fig. 4a is
remarkable, including the frequency scale: The experimental transit frequency
$f_t=V_{dc}/BWl$, marked by the arrow in Fig. 4a, was obtained by a direct
measurement of $V_{dc}$, and the maximum of the out-of-phase $r_d$ occurs at about
0.1$f_t$, in excellent agreement with the calculated behavior \cite{note2}.

Finally, we comment briefly on the response to a large $I_{ac}$ with no $I_{dc}$ as
shown in Fig. 1. In contrast to the previous situation, here all the metastable DP
exits and reenters the sample during every ac cycle, as shown previously
\cite{yosna}. The detailed analysis of this case is much more complicated since
$J_c(x,t)$ changes significantly during the ac cycle and is different for the
entering and exiting edges, resulting in highly nonlinear behavior that cannot be
treated analytically. Qualitatively, however, the maximum depth to which the DP
penetrates the sample from each edge during the corresponding half period of the ac
cycle is $x_d^{ac}=v/2f$, where $v$ is the average vortex velocity during the half
cycle, while the central part of the sample remains ordered. Therefore the fraction
of the sample occupied by the DP decreases with $f$ as $x_d^{ac}/W\propto 1/f$.
Consequently, the volume of the OP increases with frequency, the integrated $I_c$
decreases, and $V_{ac}$ increases. At sufficiently high $f$, practically the entire
sample becomes ordered so that the experimental $V_{ac}-I_{ac}$ in Fig. 1 follows
the asymptotic dashed OP curve in Fig. 3. Thus a high-frequency measurement of
$V_{ac}-I_{ac}$ is a useful method to reduce edge contamination and to approximate
the true $V-I$ of the OP.

In summary, the process of edge contamination by the metastable disordered phase in
the vicinity of the order-disorder transition is shown to explain the convex shape
of the $V-I$ characteristics in the lower part of the PE, the large difference
between $V_{dc}-I_{dc}$ and $V_{ac}-I_{ac}$ curves that grows with frequency, the
peak in the differential resistance and its rapid suppression with frequency, the
out-of-phase signal in ac transport measurements, and the opposite frequency
dependence in the different cases of ac drive.

This work was supported by the Israel Science Foundation - Center of Excellence
Program and by Minerva Foundation, Germany. E.Z. acknowledges support by the
Fundacion Antorchas - WIS Collaboration Program, and D.E.F. by RFBR grant No.
00-02-17763 and by US DOE Office of Science, contract W31-109-ENG-38.

\newpage

\ FIGURE CAPTIONS

Fig. 1. $V_{ac}-I_{ac}$ characteristics at various frequencies and $V_{dc}-I_{dc}$
($\bullet$). The voltage response increases with ac frequency.

Fig. 2. In-phase and out-of-phase differential resistance $r_d$ at dc ($\bullet$)
and various ac frequencies (left axis), and dc $V-I$ characteristic (right axis).
In-phase $r_d$ decreases with frequency, while the out-of-phase $r_d$ is maximal at
intermediate frequencies.

Fig. 3. Theoretical dc $V-I$ characteristic (solid line) with $I_c^{ord}=2$ mA,
$I_c^{dis}=18$ mA, $r_f=2$ m$\Omega$, $W=400$ $\mu$m, and
$L_r(V)=L_0(V_0/V)^{\eta}$ with $\eta =2$, $L_0=200$ $\mu$m, and $V_0=30$ $\mu$V.
The frequency dependence of the differential resistance at the operating point
($\bullet$) is shown in Fig. 4b. Inset: Schematic sample geometry and $J_c(x)$
across the sample width. $J_c(x)$ decreases with increasing current.

Fig. 4. (a) Frequency dependence of the in-phase ($\bullet$) and out-of-phase
($\circ$) $r_d$ at $I_{dc}=22$ mA near the peak of $r_d$ in Fig. 2. (b) Calculated
$r_d/r_f$ vs. $f/f_t$ at the operating point in Fig. 3.


\begin{references}
\bibitem{note2} Note that in Fig. 4 the characteristic timescale is ten times longer than the
transit time. This is one of the interesting consequences of Eq. \ref{2} and arises
from the fact that for a full rearrangement of the $J_c(x)$ distribution from
minimum to maximum (see Fig. 3 inset) the vortices have to cross the sample several
times, depending on $L_r(v)$ and the operating point on the $V-I$ curve.
\end{references}
\end{document}